\documentclass[twocolumn,showpacs,preprintnumbers]{revtex4}%
\usepackage{amsmath}
\usepackage{graphicx}
\usepackage{dcolumn}
\usepackage{bm}
\usepackage{amsfonts}
\usepackage{amssymb}%
\providecommand{\U}[1]{\protect\rule{.1in}{.1in}}
\providecommand{\U}[1]{\protect\rule{.1in}{.1in}}
\providecommand{\U}[1]{\protect\rule{.1in}{.1in}}
\providecommand{\U}[1]{\protect\rule{.1in}{.1in}}
\providecommand{\U}[1]{\protect\rule{.1in}{.1in}}
\providecommand{\U}[1]{\protect\rule{.1in}{.1in}}
\providecommand{\U}[1]{\protect\rule{.1in}{.1in}}
\providecommand{\U}[1]{\protect\rule{.1in}{.1in}}
\providecommand{\U}[1]{\protect\rule{.1in}{.1in}}
\providecommand{\U}[1]{\protect\rule{.1in}{.1in}}
\providecommand{\U}[1]{\protect\rule{.1in}{.1in}}
\providecommand{\U}[1]{\protect\rule{.1in}{.1in}}
\providecommand{\U}[1]{\protect\rule{.1in}{.1in}}
\providecommand{\U}[1]{\protect\rule{.1in}{.1in}}
\providecommand{\U}[1]{\protect\rule{.1in}{.1in}}
\providecommand{\U}[1]{\protect\rule{.1in}{.1in}}
\providecommand{\U}[1]{\protect\rule{.1in}{.1in}}
\begin{document}
\title{Achieving ground-state polar molecular condensates by chainwise atom-molecule
adiabatic passage }
\author{Jing Qian$^{1,2}$, Weiping Zhang$^{2}$, and Hong Y. Ling$^{1}$}
\affiliation{$^{1}$Department of Physics and Astronomy, Rowan University, Glassboro, New
Jersey 08028-1700, USA}
\affiliation{$^{2}$State Key Laboratory of Precision Spectroscopy, Department of Physics,
East China Normal University, Shanghai 200062, People's Republic of China}

\begin{abstract}
We generalize the idea of chainwise stimulated Raman adiabatic passage
(STIRAP) [Kuznetsova \textit{et al.} Phys. Rev. A \textbf{78}, 021402(R)
(2008)] to a photoassociation-based\ chainwise atom-molecule system, with the
goal of directly converting two-species atomic Bose-Einstein condensates (BEC)
into a ground polar molecular BEC. We pay particular attention to the
intermediate Raman laser fields, a control knob inaccessible to the usual
three-level model. We find that an appropriate exploration of both the
intermediate laser fields and the stability property of the atom-molecule
STIRAP can greatly reduce the power demand on the photoassociation laser, a
key concern for STIRAPs starting from free atoms due to the small
Franck-Condon factor in the free-bound transition.

\end{abstract}
\date{\today }

\pacs{03.75.Mn, 05.30.Jp, 32.80.Qk}
\maketitle

\section{Introduction}

A condensate of ground polar molecules with large permanent electric dipoles
represents a novel state of matter with long-range and anisotropic
dipole-dipole interactions that are highly amenable to the manipulation by dc
and ac microwave fields \cite{zoller08am}.\ As such, creation of such a
condensate is expected to be celebrated as another milestone that promises to
greatly spur activities at the forefront of physics research, particularly
with respect to quantum computing and simulation \cite{demille02} and
precision measurement \cite{precisionmeasurement}.

The road to molecular condensation is, however, complicated by the fact that
more degrees of freedom are needed to describe molecules than atoms. In
particular, cooling particles by entropy removal, a direct method popular with
atoms, has so far proved to be unable to lower the temperature of molecules
down to the regime of quantum degeneracy. Thus, most current experimental
efforts in both homonuclear \cite{homonuclear} and heteronuclear
\cite{Jin08,KKNi08,Ospelkaus08,Demille04,Demille05} molecules have all taken a
different approach exemplified by the first experimental realization of ground
polar RbCs molecules \cite{Demille05}, in which molecules are first coherently
created from ultracold atoms by photoassociation (PA) \cite{Julienne06}, and
are then brought down to the lower energy state by a coherent laser field
(instead of by spontaneous decay \cite{Demille04,Deiglmayr08,Wang04}).\ More
recently, by applying a single-step stimulated Raman adiabatic passage
(STIRAP)\cite{Bergmann98} onto the weakly bound Feshbach molecules, groups at
JILA \cite{Jin08,KKNi08,Ospelkaus08} have successfully created an ultracold
dense gas of polar $^{40}$K$^{87}$Rb molecules.

In such schemes, there is a relatively large energy difference between the
initial Feshbach and final ground molecular states. \ The former, being close
to the dissociation limit, is a highly delocalized state, while the latter is
a tightly bound state. \ It is then, in principle, difficult to locate a
single excited state, capable of a large spatial overlap integral [or
equivalently a good Franck-Condon (FC) factor] with both the initial and final
states. \ The desire to overcome this obstacle has led to the idea of
\emph{stepwise} STIRAP \cite{Jaksch02,Shapiro07}, and more recently to the
idea of \emph{chainwise} STIRAP \cite{Yelin08}, both of which are based on
models where additional intermediate states and Raman laser fields are
introduced to form a chain of $\Lambda$ systems \cite{multilevel,Vitanov98}.
\ In these STIRAPs, the two lower states within each sub-$\Lambda$ system are
far closer in energy than the initial and final states, thereby greatly
boosting the chance of locating an excited state capable of a large FC
transition to both lower states. \ In contrast to a stepwise STIRAP, which
employs a \emph{series} of STIRAPs to move molecules one step at a time down
each lower intermediate state, a chainwise STIRAP applies a \emph{single}
STIRAP between the initial and final lasers to transfer molecules, while uses
relatively high intermediate cw lasers to keep all the lower intermediate
states virtually unoccupied. \ This latter method eliminates the opportunities
for molecules to inherit decoherence from the unstable lower intermediate
states, and is clearly an improvement over the former as far as the ability to
preserve the phase-space density is concerned.

\begin{figure}[ptb]%
\centering
\includegraphics[
height=1.6345in,
width=2.6005in
]%
{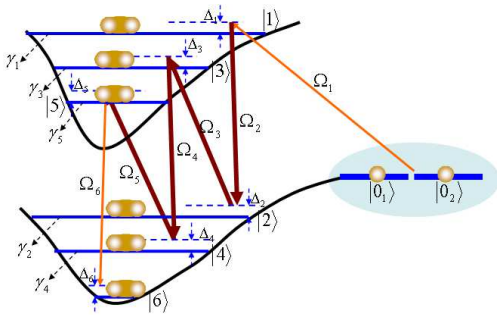}%
\caption{(Color online) A schematic of a chainwise STIRAP.}%
\label{Fig:model}%
\end{figure}

The focus of this paper is on the coupled multilevel atomic-molecular
condensate systems where the role of the initial transition is played by
photoassociation.\ (An example is provided in Fig. \ref{Fig:model}, which will
be described in detail in the next section.) \ Our goal is to develop a
generalized chainwise STIRAP founded on the concept of atom-molecule dark
state, a coherent population trapping (CPT) superposition between stable
ground species \cite{Mackie00,Ling}. \ This scheme has several attractive
properties. First, atoms are directly converted into ground molecules. \ Thus,
the loss of atoms typically associated with the initial preparation of
Feshbach molecules \cite{Jin08,Thalhammer06} is never an issue here. \ Second,
pulses of longer durations can be employed to meet the adiabatic condition; we
can do so because the atom-molecule dark state is far more stable than the
molecular dark state, where the initial state is highly unstable compared to
the ground (atom or molecule) states.\ Finally, the use of intermediate lasers
presents us with a new control knob inaccessible to typical three-level
models. It is the purpose of this paper to show that an appropriate
exploration of both the intermediate laser fields and the stability property
of the dark state can greatly reduce the power demand for the PA laser needed
for an efficient conversion. This along with other efforts involving the use
of Feshbach resonance\textbf{ }\cite{Feshbach}\textbf{ }may help to combat the
weakness in photoassociation, a key concern to STIRAPs starting from free
atoms due to the free-bound FC factor being typically very small.

Our paper is organized as follows. In Sec. \textrm{II}, we describe our model
and the underlying mean-field equations and provide the rationale that
justifies the implementation of chainwise STIRAP in our model. In Sec.
\textrm{III}, we derive a set of linearized equations around the
time-dependent CPT solution and obtain an adiabatic condition via a state
expansion over a set of orthonormal base vectors provided in Appendix B. In
Sec. \textrm{IV}, we apply this adiabatic theorem to facilitate our numerical
studies of the examples that serve to illustrate the main physics outlined in
the previous paragraph. \ Finally, a summary is given in Sec. \textrm{V}.\ \ 

\section{Model and CPT State}

Figure \ref{Fig:model} is the energy schematic diagram for a minimum model
that is capable of illustrating all the main points that we want to convey in
this paper. A laser field associates atoms from two distinct species of states
$\left\vert 0_{1}\right\rangle $ and $\left\vert 0_{2}\right\rangle $ into
molecules of state $\left\vert 1\right\rangle $ in the excited electronic
manifold with a coupling strength $\Omega_{1}$ proportional to the laser field
and the free-bound FC factor. Simultaneously, a series of laser fields of
(molecular) Rabi frequency $\Omega_{i}$ $\left(  i\geq2\right)  $ is applied
to move the molecules from the excited to the ground state $\left\vert
6\right\rangle $ via additional intermediate energy states. \ In our notation,
a molecular state $\left\vert i\right\rangle $ ($i=1,2,\cdots,6$) is coupled
to the atomic states via an $i$-photon process characterized with an
$i$-photon detuning $\Delta_{i}$ defined, respectively, as $\Delta_{1}%
=\omega_{1}-E_{1}/\hbar,$ $\Delta_{2}=\left(  \omega_{1}-\omega_{2}\right)
-E_{2}/\hbar$, $\Delta_{3}=\left(  \omega_{1}-\omega_{2}+\omega_{3}\right)
-E_{3}/\hbar,$ etc., where $\omega_{i}$ stands for the (temporal) frequency of
the laser field with Rabi frequency $\Omega_{i}$, and $E_{i}$ for the energy
of molecular state $\left\vert i\right\rangle $ relative to the free atomic
energy level. \ Further, intermediate states $\left\vert i\right\rangle $
($i=1,2,\cdots,5)$ are assumed to be unstable; for each intermediate state
$\left\vert i\right\rangle $, a decay rate $\gamma_{i}$ is introduced to
describe phenomenologically the loss of its molecules due to various
incoherent processes.

As a proof of principle, we consider, in this paper, a uniform condensate
system with a total atom number density $n$ and describe such a system with a
set of field operators $\hat{\Psi}_{i}$, where $\hat{\Psi}_{i}$ is the
operator for\ annihilating a bosonic particle in condensate state $\left\vert
i\right\rangle $. \ By following the mean-field treatment of photoassociation
at zero temperature \cite{Heinzen00} in which each $\hat{\Psi}_{i}$ is treated
as a c number $\Psi_{i}$, we obtain, from the Heisenberg's equations for
operators $\hat{\Psi}_{i}$, a set of coupled Gross-Pitaevskii's equations for
the normalized condensate fields $\psi_{i}=\Psi_{i}/\sqrt{n}$:
\begin{subequations}
\label{mean-field}%
\begin{align}
i\dot{\psi}_{0_{1}}  &  =\frac{\Omega_{1}}{2}\psi_{0_{2}}^{\ast}\psi_{1},\\
i\dot{\psi}_{0_{2}}  &  =\frac{\Omega_{1}}{2}\psi_{0_{1}}^{\ast}\psi_{1},\\
i\dot{\psi}_{1}  &  =\left(  \Delta_{1}-i\gamma_{1}\right)  \psi_{1}%
+\frac{\Omega_{1}}{2}\psi_{0_{1}}\psi_{0_{2}}+\frac{\Omega_{2}}{2}\psi_{2},\\
i\dot{\psi}_{i}  &  =\left(  \Delta_{i}-i\gamma_{i}\right)  \psi_{i}%
+\frac{\Omega_{i}}{2}\psi_{i-1}+\frac{\Omega_{i+1}}{2}\psi_{i+1}\text{,
}i=2,3,\cdots5,,\\
i\dot{\psi}_{6}  &  =\Delta_{6}\psi_{6}+\frac{\Omega_{6}}{2}\psi_{5}.
\end{align}
where the molecular Rabi frequencies
\end{subequations}
\begin{subequations}
\label{Omega}%
\begin{align}
\Omega_{1}  &  =\sqrt{2}\sqrt{n}\bar{\Omega}_{1}^{\left(  el\right)
}\left\langle v_{1}|v^{\left(  0\right)  }\right\rangle ,\\
\Omega_{i}  &  =\bar{\Omega}_{i}^{\left(  el\right)  }\left\langle
v_{i}|v_{i+1}\right\rangle ,\text{ }i=2,3,\cdots6,
\end{align}
are expressed in terms of the mean electronic Rabi frequency $\bar{\Omega}%
_{i}^{\left(  el\right)  }$, the free-bound FC factor $\left\langle
v_{1}|v^{\left(  0\right)  }\right\rangle $, and the bound-bound FC factor
$\left\langle v_{i}|v_{i+1}\right\rangle $, where $v_{i}$ and $v^{\left(
0\right)  }$ are the stationary wave functions (of interatomic distance $R$)
for a bound molecular state $\left\vert i\right\rangle $ and a pair of atoms
in states $\left\vert 0_{1}\right\rangle $ and $\left\vert 0_{2}\right\rangle
$, respectively \cite{Drummond02,Naidon03}. \ In arriving at Eqs.
(\ref{mean-field}), without the loss of the main physics, we have followed
Refs. \cite{Jin08,Yelin08} and ignored all the two-body $s$-wave
collisions.\ \ Further, in order to better illustrate the essential physics,
we will limit our study to a model in which $\Omega_{3}=\Omega_{5}\equiv
\Omega_{o}$ and $\Omega_{2}=\Omega_{4}\equiv\Omega_{e}$, where subscript $o$
and $e$ stand for the intermediate lasers of odd and even indices, respectively.

Before moving ahead, we note that Jaksch \textit{et. al.} \cite{Jaksch02} have
identified a set of rovibrational levels from $X^{1}\Sigma_{g}^{+}$ (ground)
and $A^{1}\Sigma_{u}$ (excited) electronic manifolds to implement the
homonuclear version of the model in Fig. \ref{Fig:model} for producing
$Rb_{2}$ molecules in the ground state $X^{1}\Sigma_{g}^{+}(v=0,j=0)$
[=$\left\vert 6\right\rangle $]. \ It is true that in order to arrive at a
similar set of pathways for the heteronuclear model, one must perform a
careful analysis of experimental spectroscopic data and possibly, \textit{ab
initia} calculation of various overlap integrals [FC factors defined in Eqs.
(\ref{Omega})] \cite{Deb03,Azizi04,Bigelow07}. \ However, the selection rules
for heteronuclear molecules are actually more relaxed; the heteronuclear
molecular orbitals do not have g/u symmetry, and the number of possible
transitions between the free atomic and the ground molecular state is thus
doubled. \ As a result, it is not difficult to see that such a model can be
easily generalized from homonuclear to heteronulcear molecules.

By ignoring the decays (see the justification that follows) and subjecting our
system to the conservation of total particle number,
\end{subequations}
\begin{equation}
\psi_{0_{1}}^{2}+\psi_{0_{2}}^{2}+2\sum_{i=1}^{6}\psi_{i}^{2}=1
\label{particle number conservation}%
\end{equation}
and that of atomic population difference: $\psi_{0_{1}}^{2}-\psi_{0_{2}}%
^{2}\equiv$ $0$ (or $\psi_{0_{1}}^{2}=\psi_{0_{2}}^{2}\equiv\psi_{0}^{2}$ for
a balanced model),\ we find that under the conditions of two-, four-, and
six-photon resonance, namely, $\Delta_{2}=\Delta_{4}=\Delta_{6}=0,$ the system
at steady state supports a superposition involving all the lower states with
the following amplitude distribution (see Appendix A for a detailed
derivation):
\begin{subequations}
\label{CPT}%
\begin{align}
\varphi_{0}  &  =\frac{1}{\sqrt{1+\sqrt{1+2\left(  \alpha^{2}\Omega_{1}%
/\Omega_{6}\right)  ^{2}}}},\\
\varphi_{2}  &  =-\xi\varphi_{0}^{2},\varphi_{4}=\xi\alpha\varphi_{0}%
^{2},\varphi_{6}=-\alpha^{2}\frac{\Omega_{1}}{\Omega_{6}}\varphi_{0}^{2},\\
\varphi_{1}  &  =\varphi_{3}=\varphi_{5}=0,
\end{align}
where $\alpha=\Omega_{o}/\Omega_{e}$, $\xi=\Omega_{1}/\Omega_{e}$. In arriving
at Eq. (\ref{CPT}), we have only retained the leading order term in $\xi$,
assuming that the intermediate fields $\left(  \Omega_{o},\Omega_{e}\right)  $
are far stronger than the initial and final fields $\left(  \Omega_{1}%
,\Omega_{6}\right)  $ \cite{Vitanov98,Yelin08}. (Unless stated otherwise, a
similar perturbative interpretation applies to all the other results.) \ It
needs to be stressed that Eq. (\ref{CPT}) is derived when all the decays are
ignored. \ As a result, it represents a steady state (or CPT state) of Eqs.
(\ref{mean-field}) (where all the decay rates are included) only when it does
not involve any unstable states. \ The only unstable states in Eq. (\ref{CPT})
are the lower intermediate states: $\left\vert 2\right\rangle $ and
$\left\vert 4\right\rangle $ whose amplitudes scale as $\xi$. \ Thus, it is in
the limit $\xi\ll1$ when states $\left\vert 2\right\rangle $ and $\left\vert
4\right\rangle $ remain virtually empty that this superposition can be truly
called a \textquotedblleft CPT\textquotedblright\ or \textquotedblleft
dark\textquotedblright\ state. \ Also evident is that in the same limit and
for a fixed $\alpha$, the initial and final populations are solely determined
by the ratio $\Omega_{1}/\Omega_{6}$. This lays the foundation for converting
all the atoms into the ground molecules by a chainwise STIRAP where a
counterintuitive pulse sequence is maintained only between the initial and
final laser fields. Another important feature is that the initial and final
populations are function of $\alpha^{2}\Omega_{1}$ so that the change by
$\Omega_{1}$ can also be accomplished by varying $\alpha$.\ However, the full
impact of $\alpha$ on the STIRAP has to wait until we know the adiabatic condition.\ 

\section{Adiabatic Condition}

To obtain the adiabatic condition, we linearize Eqs. (\ref{mean-field}) around
the instantaneous CPT\ state $\varphi_{i}\left(  t\right)  $ according to
$\psi_{i}\left(  t\right)  =\varphi_{i}\left(  t\right)  +\delta\psi
_{i}\left(  t\right)  $, where $\delta\psi_{i}\left(  t\right)  $ is the small
perturbation, and $\varphi_{i}\left(  t\right)  $ are Eqs. (\ref{CPT}) when
$\Omega_{i}$ are replaced with their instantaneous values $\Omega_{i}\left(
t\right)  $ at time $t$. \ This procedure results in a matrix equation for the
ket $\left\vert \delta\psi\right\rangle =\left(  \delta\psi_{0_{1}},\delta
\psi_{0_{2}},\delta\psi_{1},\cdots,\delta\psi_{6}\right)  ^{T}$:
\end{subequations}
\begin{equation}
i\frac{d}{dt}\left\vert \delta\psi\right\rangle =\left(  \mathbf{M}%
-i\mathbf{\gamma}\right)  \left\vert \delta\psi\right\rangle -i\left\vert
\dot{\varphi}\right\rangle , \label{mode equations}%
\end{equation}
where%
\begin{widetext}
\begin{equation}
\mathbf{M}=\frac{1}{2}\left(
\begin{array}
[c]{cccccccc}%
0 & 0 & \Omega_{1}^{\prime} & 0 & 0 & 0 & 0 & 0\\
0 & 0 & \Omega_{1}^{\prime} & 0 & 0 & 0 & 0 & 0\\
\Omega_{1}^{\prime} & \Omega_{1}^{\prime} & 2\Delta_{1} & \Omega_{2} & 0 & 0 &
0 & 0\\
0 & 0 & \Omega_{2} & 0 & \Omega_{3} & 0 & 0 & 0\\
0 & 0 & 0 & \Omega_{3} & 2\Delta_{3} & \Omega_{4} & 0 & 0\\
0 & 0 & 0 & 0 & \Omega_{4} & 0 & \Omega_{5} & 0\\
0 & 0 & 0 & 0 & 0 & \Omega_{5} & 2\Delta_{5} & \Omega_{6}\\
0 & 0 & 0 & 0 & 0 & 0 & \Omega_{6} & 0
\end{array}
\right)  ,\mathbf{\gamma}=\left(
\begin{array}
[c]{cccccccc}%
0 & 0 & 0 & 0 & 0 & 0 & 0 & 0\\
0 & 0 & 0 & 0 & 0 & 0 & 0 & 0\\
0 & 0 & \gamma_{1} & 0 & 0 & 0 & 0 & 0\\
0 & 0 & 0 & \gamma_{2} & 0 & 0 & 0 & 0\\
0 & 0 & 0 & 0 & \gamma_{3} & 0 & 0 & 0\\
0 & 0 & 0 & 0 & 0 & \gamma_{4} & 0 & 0\\
0 & 0 & 0 & 0 & 0 & 0 & \gamma_{5} & 0\\
0 & 0 & 0 & 0 & 0 & 0 & 0 & 0
\end{array}
\right)  ,\left\vert \dot{\varphi}\right\rangle =\left(
\begin{array}
[c]{c}%
\dot{\varphi}_{0}\\
\dot{\varphi}_{0}\\
0\\
\gamma_{2}\varphi_{2}+\dot{\varphi}_{2}\\
0\\
\gamma_{4}\varphi_{4}+\dot{\varphi}_{4}\\
0\\
\dot{\varphi}_{6}%
\end{array}
\right)  , \label{matrices}%
\end{equation}%
\end{widetext}%
with $\Omega_{1}^{\prime}=\varphi_{0}\Omega_{1}$. \ Following the standard
procedure \cite{Han}, we expand $\left\vert \delta\psi\left(  t\right)
\right\rangle =\sum c_{i}\left(  t\right)  \left\vert \lambda_{j}\left(
t\right)  \right\rangle $ in the space spanned by the instantaneous collective
modes $\left\vert \lambda_{j}\left(  t\right)  \right\rangle $ defined by
\begin{equation}
\mathbf{M}\left\vert \lambda_{j}\left(  t\right)  \right\rangle =\lambda
_{j}\left(  t\right)  \left\vert \lambda_{j}\left(  t\right)  \right\rangle .
\label{M equation}%
\end{equation}
In this new basis and under the condition that $\left\langle \lambda
_{i}\right\vert \mathbf{\dot{M}}\left\vert \lambda_{j}\right\rangle
/\left\vert \lambda_{j}-\lambda_{i}\right\vert <<1$ $\left(  \lambda_{i}%
\neq\lambda_{j}\right)  $, Eq. (\ref{mode equations}) becomes
\begin{equation}
i\frac{dc_{i}}{dt}=\lambda_{i}c_{i}-i\sum_{j}\left\langle \lambda
_{i}\right\vert \mathbf{\gamma}\left\vert \lambda_{j}\right\rangle
c_{j}-i\left\langle \lambda_{i}|\dot{\varphi}\right\rangle
\label{dressed states}%
\end{equation}
and can be put in a form convenient for us to estimate the magnitudes of
various $c_{i}$ due to the time variation of the CPT state $\left\vert
\dot{\varphi}\right\rangle $, or in another words, to arrive at the adiabatic
condition. \ (Note that whenever confusion is unlikely, we omit argument $t$
to time-dependent variables.) \ 

Thus, we see that the most crucial step in developing an adiabatic theorem is
to determine from Eq. (\ref{M equation}) a set of base vectors \ $\left\vert
\lambda_{j}\right\rangle $ upon which we can expand $\left\vert \delta
\psi\left(  t\right)  \right\rangle $. \ As this step itself can often be
quite involved, we focus on a simplified model with $\Delta_{1}=\Delta
_{3}=\Delta_{5}=0$. \ \ As we show in Appendix B, for this special model and
in the limit $\xi\ll1$, we can apply perturbation theory to obtain simple
analytical solutions from Eq. (\ref{M equation}). \ In what follows, we simply
quote the relevant results and refer interested readers to Appendix B for
details. \ In a nutshell, the collective modes are found to consist of
$\lambda_{1,2}=0$ [same as $\bar{\lambda}_{1,2}$ in Eq. (\ref{lambda1})] with
a double degeneracy, a set of \textquotedblleft soft\textquotedblright\ modes
$\lambda_{3,4}=\pm\Omega_{eff}/\left(  2\beta\right)  $ [same as $\bar
{\lambda}_{3,4}$ in Eq. (\ref{lambda1})] that scale as $\left(  \Omega
_{1}^{\prime}\text{ and }\Omega_{6}\right)  $, where $\Omega_{eff}%
=\sqrt{2\Omega_{1}^{\prime2}\alpha^{4}+\Omega_{6}^{2}}$ and $\beta
=\sqrt{1+\alpha^{2}+\alpha^{4}}$, and two sets of \textquotedblleft
stiff\textquotedblright\ modes $\lambda_{5,6}=\pm0.5\Omega_{e}\sqrt
{1+\alpha+\alpha^{2}}$ [same as $\lambda_{5,6}^{\left(  0\right)  }$ in Eq.
(\ref{lambda56})] and $\lambda_{7,8}=\pm0.5\Omega_{e}\sqrt{1-\alpha+\alpha
^{2}}$ [same as $\lambda_{7,8}^{\left(  0\right)  }$ in Eq. (\ref{lambda78})]
that scale as $\Omega_{e}$ and are larger than the \textquotedblleft
soft\textquotedblright\ modes by a factor of $\xi^{-1}$.\ Each set of
nondegenerate modes is symmetrically displaced with respect to the $0$ modes,
and the appearance of the negative modes is expected because the CPT state is
not a thermodynamical ground state.

The 0 mode being a doublet distinguishes the heteronuclear model \cite{Jing08}
from its homonuclear counterpart which only supports a single 0 mode
\cite{Han}.\ In general, due to the ground intermediate states being unstable,
the two modes here are both unstable, in contrast to the dark state in a
typical three-level system, which is a stable superposition, completely
isolated from other unstable states. However, by choosing the two modes in the
orthonormal form
\begin{subequations}
\label{lambda1_2}%
\begin{align}
\left\vert \lambda_{1}\right\rangle  &  =\frac{1}{\sqrt{2}}\left(
-1,1,0,0,0,0,0,0\right)  ^{T},\\
\left\vert \lambda_{2}\right\rangle  &  =\frac{\left(  -\Omega_{6},-\Omega
_{6},0,2\xi^{\prime}\Omega_{6},0,-2\alpha\xi^{\prime}\Omega_{6},0,2\alpha
^{2}\Omega_{1}^{\prime}\right)  ^{T}}{\sqrt{2}\Omega_{eff}},
\end{align}
where $\left\vert \lambda_{1}\right\rangle $ is exact while $\left\vert
\lambda_{2}\right\rangle $ is correct up to the first order in $\xi^{\prime
}=\Omega_{1}^{\prime}/\Omega_{e}$ which scales as $\xi\ll1$ \textbf{ }[see the
discussion bellow Eqs. (\ref{CPT})]\textbf{ }because $\xi^{\prime}=\varphi
_{0}\xi$\textbf{, }we find that $\left\vert \lambda_{1}\right\rangle $ is
completely decoupled from other states as in a true dark state while
$\left\vert \lambda_{2}\right\rangle $ couples to other states with strengths
that are at the same order of magnitude as its decay rate,
\end{subequations}
\begin{equation}
\left\langle \lambda_{2}\right\vert \mathbf{\gamma}\left\vert \lambda
_{2}\right\rangle =2\left(  \gamma_{2}+\alpha^{2}\gamma_{4}\right)
\xi^{\prime2}\Omega_{6}^{2}/\Omega_{eff}^{2}\equiv1/\tau_{0}. \label{lifetime}%
\end{equation}
The result in Eq. (\ref{lifetime}) corroborates our intuition that the use of
relatively high intermediate laser fields can indeed make the lifetime of our
CPT state, $\tau_{0}$, far longer than those of the lower intermediate states.
\ Clearly, in order to justify the use of the CPT state in Eqs. (\ref{CPT}) as
the adiabatic state for a STIRAP process, we must design the STIRAP in such a
fashion that it is slow compared with the periods of the nonzero modes but
fast compared with $\tau_{0}$, the lifetime of the CPT state. \ 

As a result, we consider all the coupling coefficients $\left\langle
\lambda_{2}\right\vert \mathbf{\gamma}\left\vert \lambda_{i}\right\rangle $
involving $\left\vert \lambda_{2}\right\rangle $ weak and ignore them from
Eqs. (\ref{dressed states}). \ In addition, we also ignore all the stiff
modes, because they are $\xi^{-1}$ times more difficult to populate than the
soft modes, whose eigenvectors are given by
\begin{equation}
\left\vert \lambda_{3,4}\right\rangle =\frac{1}{\sqrt{2}}\left(  \frac
{\alpha^{2}\Omega_{1}^{\prime}}{\Omega_{eff}},\frac{\alpha^{2}\Omega
_{1}^{\prime}}{\Omega_{eff}},\pm\frac{\alpha^{2}}{\beta},0,\mp\frac{\alpha
}{\beta},0,\pm\frac{1}{\beta},\frac{\Omega_{6}}{\Omega_{eff}}\right)  ^{T}.
\label{lambda3_4}%
\end{equation}
Under these conditions, Eqs. (\ref{dressed states}) are simplified into
\begin{subequations}
\label{soft}%
\begin{align}
\left(  i\lambda_{3}+\left\langle \lambda_{3}\right\vert \mathbf{\gamma
}\left\vert \lambda_{3}\right\rangle \right)  c_{3}+\left\langle \lambda
_{3}\right\vert \mathbf{\gamma}\left\vert \lambda_{4}\right\rangle c_{4}  &
=-\left\langle \lambda_{3}|\dot{\varphi}\right\rangle ,\\
\left\langle \lambda_{3}\right\vert \mathbf{\gamma}\left\vert \lambda
_{4}\right\rangle c_{3}+\left(  i\lambda_{4}+\left\langle \lambda
_{4}\right\vert \mathbf{\gamma}\left\vert \lambda_{4}\right\rangle \right)
c_{4}  &  =-\left\langle \lambda_{4}|\dot{\varphi}\right\rangle .
\end{align}
where we have ignored $\dot{c}_{3,4}\,$assuming that they are sufficiently
small compared to $\lambda_{3,4}$ in the adiabatic limit.

Finally, with the help of Eqs. (\ref{matrices}) and (\ref{lambda3_4}), we find
from Eqs. (\ref{soft}) that to a good approximation, the adiabatic parameter
\cite{Han}
\end{subequations}
\begin{equation}
r=\sqrt{\left\vert c_{3}\right\vert ^{2}+\left\vert c_{4}\right\vert ^{2}}/2
\label{r definition}%
\end{equation}
can be estimated according to
\begin{equation}
r=\frac{\sqrt{4\gamma_{eff}^{2}+\lambda^{2}}\alpha^{2}\dot{\chi}/\sqrt{2}%
}{\sqrt{2}\lambda^{2}\left(  1+2\alpha^{4}\chi^{2}\right)  ^{1/4}\left(
1+\sqrt{1+2\alpha^{4}\chi^{2}}\right)  }\ll1, \label{adiabaticity}%
\end{equation}
where $\chi=\Omega_{1}/\Omega_{6}$, $\lambda=$ $\Omega_{eff}/\left(
2\beta\right)  ,$ and $\gamma_{eff}=\left(  \alpha^{4}\gamma_{1}+\alpha
^{2}\gamma_{3}+\gamma_{5}\right)  /\left(  2\beta^{2}\right)  $.

\section{Discussion}

In what follows, we seek to gain from the $r$ value in Eq. (\ref{adiabaticity}%
) insights into the parameters, especially, $\alpha$, that optimize the final
conversion efficiency. In a dynamical process where the $r$ value varies with
time, we find the $r$ value evaluated at time $t_{s}$ to be a good figure of
merit that distinguishes different STIRAPs, where\ $t_{s}$ is the time when
about 50\% atoms would be converted into molecules if the system were to
follow the CPT state.\ With $t_{s}$ defined above and the Gaussian pulses
defined below, we find
\[
t_{s}=\frac{T^{2}}{2\left(  t_{1}-t_{6}\right)  }\ln\left(  \frac{2\Omega
_{6}^{0}}{\alpha^{2}\Omega_{1}^{0}}\right)  +\frac{t_{1}+t_{6}}{2},
\]
where $T$, $t_{1,6}$, and $\Omega_{1,6}^{0}$ are, respectively, the width,
peak times, and peak strengths of the Gaussian pulses: $\Omega_{1,6}\left(
t\right)  =\Omega_{1,6}^{0}\exp[-\left(  t-t_{1,6}\right)  ^{2}/T^{2}]$. In
all the calculations, $\gamma_{1,3,5}=10^{7}$ s$^{-1}$, $\gamma_{2,4}=10^{4}$
s$^{-1},$ $t_{1}=2.5T,$ $t_{6}=1.0T$, $\Omega_{1}^{0}=3.3\times10^{6}$
s$^{-1},$ $\Omega_{e}=2\times10^{8}$ s$^{-1}$, and $\Omega_{6}^{0}%
=8\times10^{7}$ s$^{-1}$. Note that although in theory high adiabaticity can
always be gained at the expense of a large $\Omega_{1}^{0}$, in practice,
$\Omega_{1}^{0}$ is quite limited due to the relative weakness in
photoassociation. For this reason, we have chosen $\Omega_{1}^{0}$ to be much
weaker than $\Omega_{6}^{0}$ ($\Omega_{1}^{0}\approx\Omega_{6}^{0}/24$). \ At
such $\Omega_{1}^{0}$, we find, using $n=10^{20}m^{-3}$ and $\mu
=4.48\times10^{-26}kg$ (reduced mass of $K$ and $Rb$ atoms), that $\Omega
_{1}^{0}/\omega_{\rho}=131,$ where $\omega_{\rho}=\hbar n^{2/3}/2\mu$
\cite{Javanainen02}. \ In a STIRAP process,\ due to quantum interference, the
molecular population in state $\left\vert 1\right\rangle $ remains extremely
small so that even when $\Omega_{1}^{0}/\omega_{\rho}$ is in the order of
10$^{2}$, rogue photodissociation of molecules in state $\left\vert
1\right\rangle $ is shown to produce a negligible fraction of noncondensate
atom pairs \cite{Mackie04}. \ As a result, we ignore the rogue
photodissociation in this work.%
\begin{figure}[ptb]%
\centering
\includegraphics[
height=3.5898in,
width=2.6801in
]%
{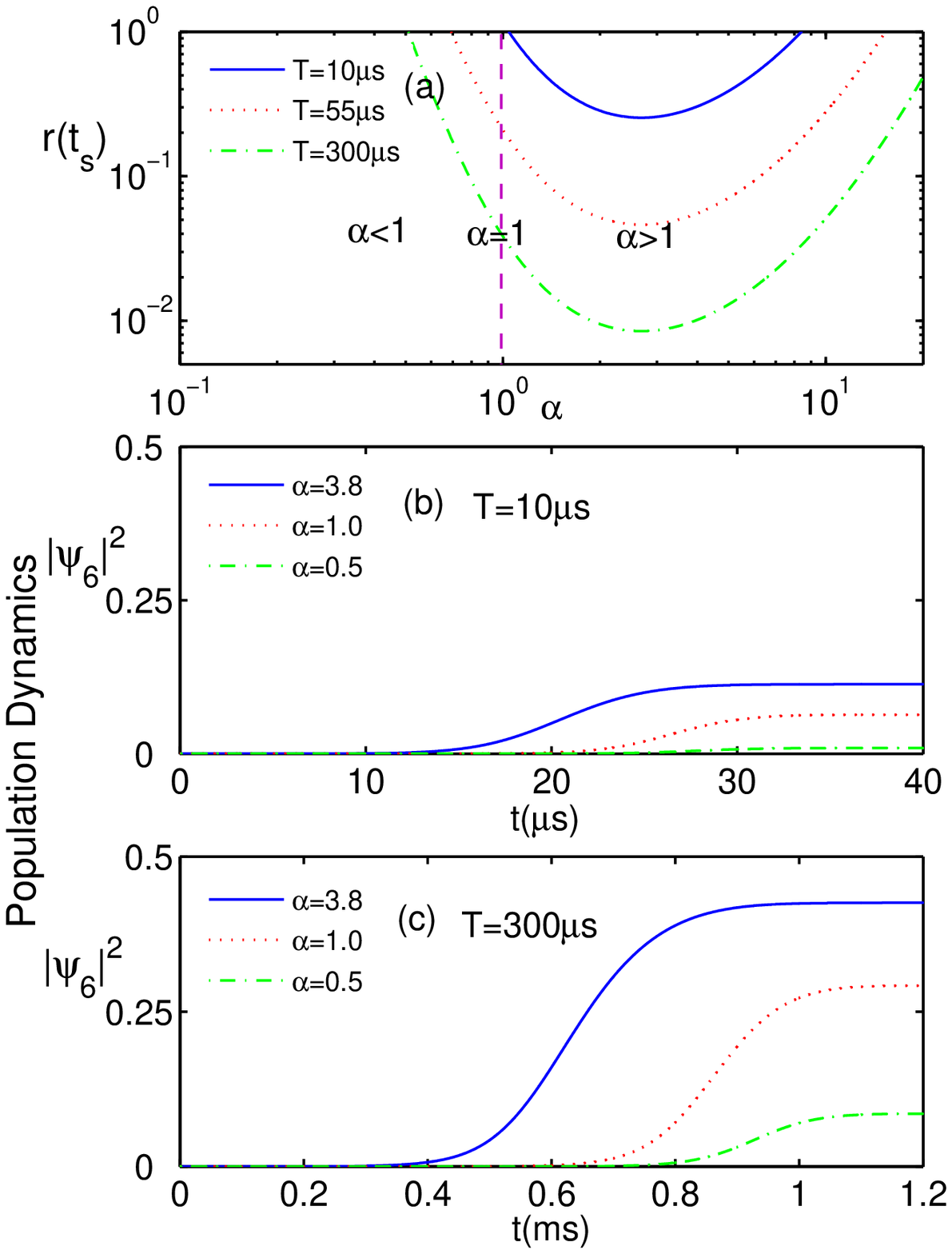}%
\caption{(Color online) (a) The $r$ value at $t_{s}$ as a function of $\alpha$
under different $T$.\ The population dynamics with (b) $T=10$ $\mu s$ and (c)
$T=300$ $\mu s$ for different $\alpha$.\ Other parameters are defined in the
text.}%
\label{Fig:comparision}%
\end{figure}
\ 

Figure \ref{Fig:comparision}(a) illustrates how the $r$ value at $t_{s}$
changes with $\alpha$ under different $T$.\ The most interesting feature here
is that, for a given $T$, the $r$ value is quite high on the side of
$\alpha<1$ (the left side of the dashed vertical line $\alpha=1$)$,$\ but
significantly smaller within certain region on the side of $\alpha>1$. As a
result, we see in Fig. \ref{Fig:comparision}(b) with $T=10$ $\mu s$ that when
$\alpha$ changes from 0.5 to 3.8 (at which, $r$ value is near minimum), the
conversion efficiency $\eta\lbrack=2\psi_{6}^{2}\left(  \infty\right)
]$\ increases from 1.8\% to 22.6\%, a trend consistent with Fig.
\ref{Fig:comparision}(a) with $T=10$ $\mu s$. But, we caution that no
efficiencies significantly higher than 22.6\% are possible in this case by
further raising the $\alpha$ value because the $r$ value actually increases
with $\alpha$ when $\alpha$ is sufficiently large according to Fig.
\ref{Fig:comparision}(a); we trace this to the fact that unlike the CPT\ state
in Eq. (\ref{CPT}), $\lambda$ and $\gamma_{eff}$ and hence the adiabatic
condition in Eq. (\ref{adiabaticity}) do not scale as $\alpha^{2}\Omega_{1}$.

An important point to make is that were the lifetime of the dark state limited
to the order of $10$ $\mu s$, a higher efficiency would indeed have to come at
the expense of a higher PA laser power. This, however, is not needed owing to
another important virtue of our system - the stability of our CPT state, whose
lifetime can be made much longer than those of the lower states in the ground
electronic manifold. This, therefore, affords us with a plenty of room to
increase efficiency\ by using pulses with longer durations rather than higher
powers. Indeed, the population dynamics in Fig. \ref{Fig:comparision}(c) with
a much longer pulse ($T=300$ $\mu$s$)$ demonstrates, in addition to a trend
same as in Fig. \ref{Fig:comparision}(b), a dramatic increase in the maximum
efficiency, which can now reach more than 85\%.

To estimate the required power on the photoassociation laser field, we
consider heteronuclear molecules involving $Rb$ atoms, for example, $KRb$
\cite{Wang04} and $RbCs$ \cite{Demille05}. \ The free-bound FC factor for
heteronuclear molecules is expected to be smaller than that for their
homonuclear counterparts because the excited potential at large internuclear
distance $R$ for the heteronuclear molecules is dominated by the van der Waals
potential (R$^{-6}$) and thus has a shorter range than that for the
homonuclear molecules, which is dominated by the resonant dipole-dipole
interaction (R$^{-3}$). \ An encouraging news according to \textit{ab initia}
calculations in Refs. \cite{Azizi04,Ghosal09} is that the former is only
slightly smaller than the latter. \ As a result, in our estimation, we choose
a free-bound FC factor $\left\langle v_{1}|v^{\left(  0\right)  }\right\rangle
=3\times10^{-14}$ m$^{3/2}$ several times smaller than that of $Rb_{2}$
molecules, which can be on the order of 10$^{-13}$m$^{3/2}$ (for $n=10^{20}$
m$^{-3}$) according to Naidon and Masnou-Seeuws \cite{Naidon03}. \ Finally,
using 1.6 $ea_{0}$ as the atomic dipole moment \cite{Tiesinga03} with $e$ the
electron charge and $a_{0}$ the Bohr radius, we estimate the peak PA laser
intensity to be $3$.$8\times10^{3}$ W/cm$^{2}$. \ This admittedly high (and
yet attainable) intensity can be put in perspective by comparison with the
case where $T=10$ $\mu$s and $\alpha=1$, where an intensity of more than $100$
times higher is needed to achieve the same level of high efficiency [in Fig.
\ref{Fig:comparision}(c) with $\alpha=3.8$].

\section{Summary}

In summary, we have generalized the chainwise STIRAP from pure molecular to
coupled atomic-molecular systems.\ In addition to the known advantages, for
example, the increased chance to locate pairs of Raman transitions with large
FC factors, we have uncovered additional virtues.\ In particular,\ $\alpha$, a
ratio between intermediate laser fields, was found to serve as a robust
experimental control knob, inaccessible to the usual three-level systems. This
control knob together with the stability of the atom-molecule dark state may
bring us one step closer to overcome the PA weakness, so that the ground polar
molecules can be created directly from degenerate atomic gases in a manner
that preserves the phase-space density.

\section{Acknowledgement}

This work is supported by the US National Science Foundation (H.Y.L), the U.S.
Army Research Office (H.Y.L.), and the National Natural Science Foundation of
China under Grant No. 10588402, the National Basic Research Program of China
(973 Program) under Grant No. 2006CB921104, the Program of Shanghai Subject
Chief Scientist under Grant No. 08XD14017, and the Program for Changjiang
Scholars and Innovative Research Team in University, Shanghai Leading Academic
Discipline Project under Grant No. B480 (W.Z.).

\appendix

\section{}

This appendix provides the steps that we take to arrive at Eqs. (\ref{CPT}).
We begin with Eqs. (\ref{mean-field}) at steady state, where the derivatives
on the left-hand sides are all set to zero. \ Next, we ignore all the
decays\ as well as all the excited populations $\left(  \psi_{1}=\psi_{3}%
=\psi_{5}=0\right)  $. \ We then see that the equations for $\psi_{2},\psi
_{4}$, and $\psi_{6}$ lead to the CPT\ condition: $\Delta_{2}=\Delta
_{4}=\Delta_{6}=0$, while the rest of equations are simplified to
\begin{subequations}
\label{simplfied}%
\begin{align}
-\xi\psi_{0_{1}}\psi_{0_{2}}  &  =\psi_{2},\\
-\alpha\psi_{2}  &  =\psi_{4},\\
-\frac{\Omega_{o}}{\Omega_{6}}\psi_{4}  &  =\psi_{6}.
\end{align}
where as in the main text we have made the use of $\Omega_{2}=\Omega
_{4}=\Omega_{e},$ $\Omega_{3}=\Omega_{5}=\Omega_{o}$, $\alpha=\Omega
_{o}/\Omega_{e}$ and $\xi=\Omega_{1}/\Omega_{e}$. \ For a balanced system with
$\psi_{0_{1}}^{2}=\psi_{0_{2}}^{2}\equiv\psi_{0}^{2}$, we find from Eqs.
(\ref{simplfied}) that $\psi_{2}=-\xi\psi_{0}^{2}$, $\psi_{4}=\alpha\xi
\psi_{0}^{2}$, and $\psi_{6}=-\left(  \Omega_{o}/\Omega_{6}\right)  \alpha
\xi\psi_{0}^{2}$, which, when combined with the particle number conservation
in Eq. (\ref{particle number conservation}), gives rise to
\end{subequations}
\begin{align*}
\varphi_{0}^{2}  &  =\frac{1}{1+\sqrt{1+2\left(  1+\alpha^{2}\right)  \xi
^{2}+2\left(  \alpha^{2}\Omega_{1}/\Omega_{6}\right)  ^{2}}},\\
\varphi_{2}  &  =-\xi\varphi_{0}^{2},\varphi_{4}=\xi\alpha\varphi_{0}%
^{2},\varphi_{6}=-\alpha^{2}\frac{\Omega_{1}}{\Omega_{6}}\varphi_{0}^{2}.
\end{align*}
\ \ Clearly, we see that up to the first order in $\xi$, Eqs. (\ref{simplfied}%
) become Eqs. (\ref{CPT}) in the main text.

\section{}

In this appendix, we show how to obtain from Eq. (\ref{M equation}) the
eigenvalues and eigenvectors (that are needed to derive the adiabatic
condition) in the limit of $\xi\ll1$ for the special case of $\Delta
_{1,3,5}=0$. \ To begin, we divide $\mathbf{M}$ in Eq. (\ref{matrices}) into
two parts:%
\begin{equation}
\mathbf{M=M}_{0}+\mathbf{M}^{\prime}%
\end{equation}
where%
\begin{widetext}
\begin{equation}
\mathbf{M}_{0}=\frac{1}{2}\left(
\begin{array}
[c]{cccccccc}%
0 & 0 & 0 & 0 & 0 & 0 & 0 & 0\\
0 & 0 & 0 & 0 & 0 & 0 & 0 & 0\\
0 & 0 & 0 & \Omega_{e} & 0 & 0 & 0 & 0\\
0 & 0 & \Omega_{e} & 0 & \Omega_{o} & 0 & 0 & 0\\
0 & 0 & 0 & \Omega_{o} & 0 & \Omega_{e} & 0 & 0\\
0 & 0 & 0 & 0 & \Omega_{e} & 0 & \Omega_{o} & 0\\
0 & 0 & 0 & 0 & 0 & \Omega_{o} & 0 & 0\\
0 & 0 & 0 & 0 & 0 & 0 & 0 & 0
\end{array}
\right)  ,\mathbf{M}^{\prime}=\frac{1}{2}\left(
\begin{array}
[c]{cccccccc}%
0 & 0 & \Omega_{1}^{\prime} & 0 & 0 & 0 & 0 & 0\\
0 & 0 & \Omega_{1}^{\prime} & 0 & 0 & 0 & 0 & 0\\
\Omega_{1}^{\prime} & \Omega_{1}^{\prime} & 0 & 0 & 0 & 0 & 0 & 0\\
0 & 0 & 0 & 0 & 0 & 0 & 0 & 0\\
0 & 0 & 0 & 0 & 0 & 0 & 0 & 0\\
0 & 0 & 0 & 0 & 0 & 0 & 0 & 0\\
0 & 0 & 0 & 0 & 0 & 0 & 0 & \Omega_{6}\\
0 & 0 & 0 & 0 & 0 & 0 & \Omega_{6} & 0
\end{array}
\right)  . \label{M0M'}%
\end{equation}%
\end{widetext}
Here, $\mathbf{M}_{0}$ is an unperturbed part including all the intermediate
Rabi frequencies $\Omega_{o}$ and $\Omega_{e}$, while $\mathbf{M}^{\prime}$,
consisting of only the initial and final fields $\Omega_{1}^{\prime}$ and
$\Omega_{6}$, can be regarded as a perturbation to $\mathbf{M}_{0}$ in the
limit of $\xi\ll1$. \ Next, we determine from the equation
\begin{equation}
\mathbf{M}_{0}\left\vert \lambda_{j}^{\left(  0\right)  }\right\rangle
=\lambda_{j}^{\left(  0\right)  }\left\vert \lambda_{j}^{\left(  0\right)
}\right\rangle \label{M0}%
\end{equation}
eigenvalues $\lambda_{j}^{\left(  0\right)  }$ and eigenstates $\left\vert
\lambda_{j}^{\left(  0\right)  }\right\rangle $ of the unperturbed part
$\mathbf{M}_{0}$. \ We find that $\lambda_{j}^{\left(  0\right)  }$ take the
following values:%

\begin{subequations}
\label{zero lambda}%
\begin{align}
&  \lambda_{1,2,3,4}^{\left(  0\right)  }=0,\label{lambda0}\\
&  \lambda_{5,6}^{\left(  0\right)  }=\frac{\pm\Omega_{e}\sqrt{1+\alpha
+\alpha^{2}}}{2},\label{lambda56}\\
&  \lambda_{7,8}^{\left(  0\right)  }=\frac{\pm\Omega_{e}\sqrt{1-\alpha
+\alpha^{2}}}{2}. \label{lambda78}%
\end{align}
The zero eigenvalue has a four-fold degeneracy, and the corresponding
(orthonormalized) eigenstates are found from Eq. (\ref{M0}) to take the form
\end{subequations}
\begin{subequations}
\label{four degenerate states}%
\begin{align}
\left\vert \lambda_{1}^{\left(  0\right)  }\right\rangle  &
=(1,0,0,0,0,0,0,0)^{T},\\
\left\vert \lambda_{2}^{\left(  0\right)  }\right\rangle  &
=(0,1,0,0,0,0,0,0)^{T},\\
\left\vert \lambda_{3}^{\left(  0\right)  }\right\rangle  &
=(0,0,0,0,0,0,0,1)^{T},\\
\left\vert \lambda_{4}^{\left(  0\right)  }\right\rangle  &  =(0,0,\frac
{\alpha^{2}}{\beta},0,-\frac{\alpha}{\beta},0,\frac{1}{\beta},0)^{T},
\end{align}
where $\beta=\sqrt{1+\alpha^{2}+\alpha^{4}}$. \ This degeneracy, however, can
be partially lifted by the perturbation $\mathbf{M}^{\prime}$ as we will show
shortly. \ $\lambda_{5,6}^{\left(  0\right)  }$ and $\lambda_{7,8}^{\left(
0\right)  }$ are proportional to $\Omega_{e}$ and are far larger in magnitude
than those split from the zero eigenvalue by $\mathbf{M}^{\prime}$; the modes
associated with the former eigenvalues are far more difficult to populate than
those associated with the latter eigenvalues. \ Thus, it suffices, for our
purpose, that we only focus on the\ eigensubspace spanned by the four base
vectors in Eqs. (\ref{four degenerate states}), in which $\mathbf{M}^{\prime}$
in Eq. (\ref{M0M'}) has the following matrix representation
\end{subequations}
\begin{equation}
\mathbf{M}^{\prime}=\frac{1}{2\beta}\left(
\begin{array}
[c]{cccc}%
0 & 0 & 0 & \Omega_{1}^{\prime}\alpha^{2}\\
0 & 0 & 0 & \Omega_{1}^{\prime}\alpha^{2}\\
0 & 0 & 0 & \Omega_{6}\\
\Omega_{1}^{\prime}\alpha^{2} & \Omega_{1}^{\prime}\alpha^{2} & \Omega_{6} & 0
\end{array}
\right)  \label{M prime}%
\end{equation}
where the use of $\mathbf{M}_{ij}^{\prime}=\left\langle \lambda_{i}^{\left(
0\right)  }\right\vert \mathbf{M}^{\prime}\left\vert \lambda_{j}^{\left(
0\right)  }\right\rangle $ has been made. \ In the spirit of degenerate
perturbation theory \cite{book}, we form the following linear combination of
degenerate states in the four-dimensional Hilbert space%
\begin{equation}
\left\vert \bar{\lambda}_{n}\right\rangle =\sum_{i=1}^{4}a_{ni}\left\vert
\lambda_{i}^{\left(  0\right)  }\right\rangle \label{linear expansion}%
\end{equation}
where $\mathbf{a}_{n}=\left(  a_{n1},a_{n2},a_{n3},a_{n4}\right)  ^{T}$ is the
eigenvector of matrix $\mathbf{M}^{\prime}$ with an eigenvalue $\bar{\lambda
}_{n}$ or equivalently it satisfies the following equation%
\begin{equation}
\mathbf{M}^{\prime}\mathbf{a}_{n}=\bar{\lambda}_{n}\mathbf{a}_{n}
\label{M prime lambda_n}%
\end{equation}
By solving Eq. (\ref{M prime lambda_n}), we find the following set of
eigenvalues
\begin{equation}
\bar{\lambda}_{1,2}=0,\bar{\lambda}_{3,4}=\pm\frac{\Omega_{eff}}{2\beta}
\label{lambda1}%
\end{equation}
where $\Omega_{eff}=\sqrt{2\Omega_{1}^{\prime2}\alpha^{4}+\Omega_{6}^{2}}$.
\ As can be seen, $\mathbf{M}^{\prime}$ reduces the degeneracy of the zero
mode from four folds to two folds, creating a pair of so-called soft modes,
whose eigenvalues, $\bar{\lambda}_{3,4}$, are symmetrically displaced from
zero eigenvalue. \ The (unit normalized) eigenvectors $\mathbf{a}_{3,4}$ for
$\bar{\lambda}_{3,4}$ are found from Eq. (\ref{M prime lambda_n}) to take the
form
\begin{equation}
\mathbf{a}_{3,4}=\frac{1}{\sqrt{2}}\left(  \frac{\alpha^{2}\Omega_{1}^{\prime
}}{\Omega_{eff}},\frac{\alpha^{2}\Omega_{1}^{\prime}}{\Omega_{eff}}%
,\frac{\Omega_{6}}{\Omega_{eff}},\pm1\right)  ^{T},
\end{equation}
which, when combined with Eq. (\ref{linear expansion}), yields the desired
soft modes in Eq. (\ref{lambda3_4}).

At this point, we stress that the zero mode of two-fold degeneracy cannot be
lifted by $\mathbf{M}^{\prime}$ as one can easily check directly from Eq.
(\ref{M equation}) with $\lambda_{j}=0$ that it supports two linearly
independent (but nonorthogonal) solutions%
\begin{align}
\left\vert b_{1}\right\rangle  &  =\left(  -1,1,0,0,0,0,0,0\right)  ^{T},\\
\left\vert b_{2}\right\rangle  &  =\left(  -\frac{\Omega_{e}}{\Omega
_{1}^{\prime}},0,0,1,0,-\alpha,0,\frac{\alpha\Omega_{o}}{\Omega_{6}}\right)
^{T}.
\end{align}
Finally, we apply Gram-Schmidt orthogonalization to transform $\left\vert
b_{1,2}\right\rangle $ into a set of orthonormalized vectors $\left\vert
\lambda_{1,2}\right\rangle $ in Eq. (\ref{lambda1_2}).%

\end{document}